# Human-in-the-loop Machine Learning: A Macro-Micro Perspective

**Jiangtao Wang[1]**   **Bin Guo[2]**   **Luke Chen[3]**
1. Coventry University, Coventry, United Kingdom
2. Northwestern Polytechnical University, Xi'an, China
3. Ulster University, Newtownabbey, United Kingdom

*Jiangtao.wang@conventry.ac.uk, guobin.keio@gmail.com, l.chen@ulster.ac.uk*

**Abstract:** Though technical advance of artificial intelligence and machine learning has enabled many promising intelligent systems, many computing tasks are still not able to be fully accomplished by machine intelligence. Motivated by the complementary nature of human and machine intelligence, an emerging trend is to involve humans in the loop of machine learning and decision-making. In this paper, we provide a macro-micro review of human-in-the-loop machine learning. We first describe major machine learning challenges which can be addressed by human intervention in the loop. Then we examine closely the latest research and findings of introducing humans into each step of the lifecycle of machine learning. Finally, we analyze current research gaps and point out future research directions.

## 1. Introduction

Advances in machine learning (ML) technologies have led to an explosion of major breakthroughs in the field of artificial intelligence (AI), which in turn has given rise to ubiquitous intelligent systems in our daily life. Typical applications include autonomous vehicles, game playing (e.g., AlphaGo), precise medical diagnosis and prescription, assistive robots, face recognition, and so forth. Despite the increasing usage and power of ML in these AI applications, there is still a large spectrum of computing tasks that pure machine intelligence cannot fully handle due to challenges such as lack of large-scale dataset, low data quality, insufficient training labels, and explanability and reliability issues of ML-based decision-making.

Motivated by the complementary nature of human and machine intelligence [Kamar 2016] [Wang 2019], an approach that exploits human's cognitive power to help address these ML challenges has emerged recently, which is referred to as "Human-in-the-loop Machine Learning (HML)". HML is intended to integrate relevant human competences into the whole cycle of ML, ranging from data collection, algorithm tuning, parameters selection, to the usage of the outcomes of ML to actuate the physical world. The fundamental goal of this emerging topic is to re-examining and reframing the ML workflow from a human-centered perspective. The study of HML has the following significance. First, it can generate more usable and human-centric ML tools for AI engineers or even application end users. Second, it can lead to an in-depth understanding about how the power of human intelligence and machine intelligence can be combined to solve real-world computing problems. Inspired by the aforementioned visions, a number of studies for HML have been recently presented in top conference or prestigious journals, which has become a trendy research topic across multiple overlapping research communities (e.g., AI, Human Computer Interaction, crowd and social computing, data mining, and so on).

There are several literature reviews or tutorials summarizing the state of the art with each focusing on different topical area, see Table 1. Nevertheless, they only touch individual specific research problem within the scale of HML. To the best of our knowledge, there are no survey papers to present a global picture about how human cognitive power can be appropriately integrate into ML processes. A systematic study and classification of the research problems in the HML research domain is still missing.

To this end, we attempt to fill the gap by systematically studying and classifying the research work of HML from a *macro-micro* perspective. Instead of discussing a particular aspect such as interface design [Dudley 2018] or the explainable models [Riccardo 2019], this paper is to provide a high-level overview of HML for those researchers who want to step into this topic area for the first time. More specifically, this paper makes the following contributions in this paper.

- From a *macro* perspective of the complementary nature of machine and human, we identify the main bottlenecks of ML, and then provide deep insights about why and how human can collaborate with machine so as to address these challenges.

- We present a *macro-micro* review by dividing the representative studies into several stages of the ML lifecycle. Here, in order to help the readers to establish a clearer and more understandable picture, the reviews are organized in our proposed structure/framework called Human4ML.

- We analyze existing research gaps in this field with a number of future research directions and proposals, which may contribute to new ideas and visions.

It should be noted that this paper does not aim to give a comprehensive review of HML due to the page limit. We provide a general framework into which other existing works can be easily added by the readers themselves if they intend to extend it to a more comprehensive one.

Table 1: Relevant Reviews/Tutorials and Their Focused Aspects

| Relevant Survey | Topical Areas |
| --- | --- |
| [Dudley 2018] | Interface design techniques in the interactive machine learning |
| [Riccardo 2019] | Models and approaches to make AI more explainable |
| [Zhang 2018] | Visualization approaches for interpreting deep-learning-based models |
| [Wang 2019] | Optimization for crowdsourcing and ML (e.g., labeling and inference) |
| [Zhang 2019] | Leveraging Human Guidance for Deep Reinforcement Learning Tasks |

## 2. The Needs for Human Involvement

Although ML is powerful in developing a variety of smart systems, it still encounters a number of inevitable challenges under real-world practical situations. While these challenges can be addressed in many different ways, in this paper we focus on those challenges that can be potentially alleviated by introducing human intelligence into the process of ML.

**Challenge 1: Dirty and Incomplete Data.**

The most fundamental ingredient of an AI system is data for training the learning model, because ML is essentially the process of automatically learning knowledge from data. The quality of data is crucially important for the performance of a learning model. Unfortunately, datasets acquired in the real world can be low quality. Dirty or incomplete data, such as missing values, typos, inconsistent values, out-of-date records, usually leads to inaccurate analysis and learning. Several automatic data cleaning approaches have been developed to enhance data quality, but their performance is still far from meeting the requirement of applications [Xu 2016]. In many scenarios, we still need the help from human workers to detect, repair or eliminate dirty or incomplete data.

**Challenge 2: Heterogeneity and Isolation**

Many intelligent applications need data feed from multiple domains, which are usually originated from heterogenous and isolated ICT systems. For example, if we attempt to develop an AI-based system to predict population health status in different regions/countries in EU, the first challenge is data integrations and linkage from different information systems to build a comprehensive training dataset for model training. However, this task is non-trivial because the data structure and database design might be different for systems with similar semantics, which increase the difficulty and workload of data linkage. For example, if we want to get an overall picture of heart attacks in Europe, we need to integrate and link relevant data from heterogenous information system of many hospitals from multiple countries. Although a lot of data integration tools such as schema matching have been proposed and proved to be effective to some extent, they are often uncertain about schema matchings they suggest, and this uncertainty is inherent since it arises from the inability of the schema to fully capture the semantics of the represented data [Zhang 2018]. Human common sense can often help in this process.

**Challenge 3: Insufficient Labels in Training Data**

In recent years, the real-world impact of ML has grown in leaps and bounds. However, ML models heavily rely on massive sets of hand-labeled training data, and the cost of building such labeled datasets is one of the key bottlenecks to large-scale ML adoption. Although some weak supervised learning approaches [Zhou 2017] can ease this problem in some cases, it is still preferable in general if we can obtain sufficient labels for training. Since labeling a large amount of data is time-consuming, there is an emerging trend to leverage the power of the crowdsourcing workers to accomplish this task.

**Challenge 4: Feature Construction with Limited Data**

Constructing feature set is an essential but uneasy step to ML. There are two common approaches for feature discovery or extraction: the automatic methods, e.g., deep learning and domain-knowledge-based methods (e.g., feature engineering). However, if we are only given a limited amount of training data, both approaches become ineffective as irrelevant features extracted from limited datasets can lead to the problem called overfitting. For example, it is easy to distinguish a dog from a cat in an image with deep learning as the relevant labeled training data on the Web is huge. However, the classification between "Northern Flicker" and "Red Bellied Woodpecker" [Deng 2016]) is much more challenging when there are insufficient training data and fewer discriminative features.

**Challenge 5: Trust and Transparency**

ML-based AI systems sometimes are completely treated as black boxes without human supervision, where the learning models sometimes learn undesirable tricks that do an optimal job of satisfying pre-programmed goals (e.g. minimizing loss function) on the training data without reflecting the complicated implicit desires of the human system designers. In some ML-enabled applications, the black box issue doesn't matter if the task is not very critical. However, this issue becomes very sensitive when the application domain is critical (e.g., medical treatment, and security/safety control), where knowing the reasons behind the machine's decision is of high importance. Thus, it is of vital significance to open that black box to achieve a more transparent, explainable, and trustworthy AI and ML with human in the loop [Riccardo 2019].

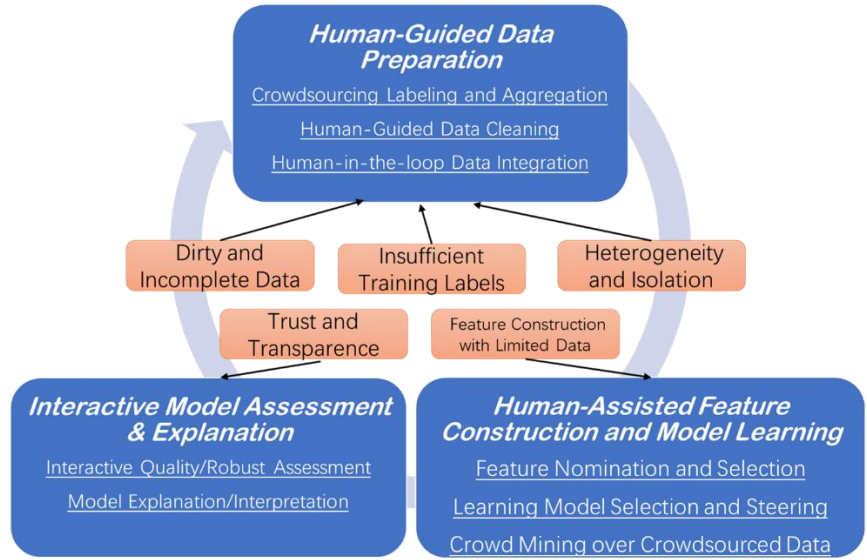

Fig 1. Human4ML framework: organizing representative studies

## 3. Structure Overview: Lifecycle Perspective

The typical ML process consists of three iterative stages: data pre-preparing (e.g., data collection and pre-processing), feature construction and model learning, model assessment. Human intelligence can be integrated into each of the three stages. To this end, we present an overview of the state-of-the-art HML research using a general framework named Human4ML following the ML lifecycle, as shown in Fig.1. In this framework, corresponding human-computation-based techniques are adopted to improve certain stage or task of ML. From Figure 1, we can see an overview of the utilization of human capabilities in each stage of ML. Here, we characterize the features of each stage by using the logic of "2W1H" (What, Who, and How) in Table 2. "What" refers to what problems or challenges each component aims to support, "How" refers to the general idea to achieve the goal, and "who" refers to the stakeholders who are involved in this stage.

### Phase 1: Human-Guided Data Preparation

Human intelligence can be leveraged in data collection and pre-processing procedures (we call them together as data preparation in this paper) to address the challenge of low quality and insufficiency of training dataset mentioned above. To our best knowledge, human computation techniques (e.g., crowdsourcing) has primarily adopted in three tasks in data preparation phase, including data labeling and aggregation [Liu 2018], data cleaning [Xu 2016], and data integration [Li 2017]. Some representative research works related to this component will be introduced in the Section 4 with more details.

### Phase 2: Human-Assisted Feature Construction and Model Learning

For either automatic or domain expert-based approaches, feature construction and model learning is still a challenge for fine-grained features with limited training data. Recent studies show the potential in inviting human (either experts or crowdsourcing workers) to collaborate with machine algorithms to better elicit good features and perform model training. The key insights lie in two aspects. One is that human workers are better at nominating subtle behavioral signals than machines [Cheng 2015], so that they can complement machine intelligence in feature extraction. The other is that the selection of learning models are significantly correlated with training data and learning target, and human can help compare, select, and steer different ML models (e.g., choosing the right classifiers in the classification task) [Das 2019]. We will introduce related work for this component in Section 5.

Table 2. Using "2W1H" to characterize Human4ML Framework

| Stages | WHAT | HOW | WHO |
| --- | --- | --- | --- |
| *Human-guided Data Preparation* | To obtain more labels, clean data, and integrate data (address challenge 1,2 and 3) | Use crowdsourcing or invite domain experts to help | Workers in crowdsourcing platform, domain experts, source data managers |
| *Human-Assisted Feature Construction and Model Learning* | Collaborate with machine to construct feature set and perform model training (address challenge 4) | Nominate/select features (based on crowdsourcing), select/steer models by designing suitable visualization interface, and perform crowd-based data mining over crowdsourced data | AI engineers, workers in crowdsourcing platforms, domain experts |
| *Interactive Model Assessment and Explanation* | Quality assurance, present the reasons behind the decision of ML models (address challenge 5) | Design human-computer interface, visualization, adopt explainable models | AI engineers, domain experts |

**Phase 3: Interactive Model Assessment and Explanation**

Only when we have adequate trust, can we use a ML model in real world, especially for critical systems (e.g., medical, military, financial, and so on). Although good performance in simple indicators on testing sets (e.g., precision, recall, etc.) can provide some belief to some extent, human themselves, also serving as the end user of the ML applications, cannot be excluded out of the evaluation process. There are two key research issues human can actively participate in this stage, namely (1) enabling quality and robustness assessment with human in the loop, and (2) providing model interpretability by designing effective human-computer interfaces to bridge the gap between machine and human. Representative studies for this component will be introduced in Section 6.

## 4. Human-Guided Data Preparation

In this paper, the data preparation consists of key operations such as data collection, data cleaning, and data integration. Accordingly, this section introduces how human intelligence is utilized in these operations respectively.

### 4.1 Crowdsourcing Labeling and Aggregation

In the data preparing phase, the most studied task is to use crowdsourcing [Howe 2006] in the data labeling and aggregation. The naïve way for label aggregation is majority vote, where the aggregated value is regarded as the opinion of the majority of the crowd workers. However, in realistic scenarios, workers are heterogenous in skills, expertise, and reputation, and so on. Therefore, weight-based aggregation methods are proposed, where the label of a worker with high weight would count more. Aggregation methods that consider the diversity and dependency of workers are proposed in recent years to improve the label aggregation accuracy [Nushi 2015] [Venanzi 2016]. In addition to the aggregation approach, worker selection and task assignment is also important. For example, authors in [Liu 2018] studied in an active learning setting where training samples are adaptively selected to be labeled, and a learner can query a set of crowdsourcing workers with unknown expertise level for label information.

**4.2 Human-Guided Data Cleaning**

Detecting and repairing dirty data is one of the perennial challenges in data analytics, and failure to do so can result in inaccurate analytics and unreliable decisions. In the data preparation phase, another task that human workers can help is data cleaning. For example, the authors present CrowdCleaner in [Tong 2014], a smart system to clean multi-version data on the Web. In this system, the authors leverage active and passive crowdsourcing methods together for rectifying errors for multi-version data. In another example, [Chu 2015], the authors use both human computation and open knowledge base to generate top-k possible repairs for incorrect data. More extensive review for this technical aspect can be referred to [Xu 2016].

**4.3 Human-in-the-loop Data Integration**

Human intelligence and crowdsourcing techniques are extensively used in many data integration tasks in recent years.

(1) *Schema matching in Database*. Schema matching is a central challenge for data integration systems. Automated tools are often uncertain about schema matchings they suggest, and this uncertainty is inherent since it arises from the inability of the schema to fully capture the semantics of the represented data. Inspired by the popularity and the success of easily accessible crowdsourcing platforms, a number of studies such as [Li 2017] explore the technique of human computation and crowdsourcing to reduce the uncertainty of schema matching with the consideration of cost and quality. (2) *Entity linkage in Knowledge Base/Graph*. In entity linking, words of interest (names of persons, locations and companies) are mapped from an input text to corresponding unique entities in a target knowledge base (KB), which serves as a basic component of many ML and AI applications. Several automatic linking methods have been proposed. However, due to the inconsistency and uncertainty of large-scale KBs, automatic techniques for entity linkage achieve low quality. Thanks to the open crowdsourcing platforms, studies such as [Zhuang 2017] propose solutions to harness the human workers to improve the linkage quality.

# 5. Human-Assisted Feature Construction and Model Learning

Human intelligence can be used to nominate and select features in ML, and further help in model selection and steering.

**5.1 Feature Nomination and Selection**

Crowdsourcing and human computation can be utilized to discover (nominate or select) features for supervised/unsupervised learning. For example, the study in [Cheng 2015] presented a hybrid crowd-machine learning framework called Flock. The basic idea of the framework is that: it starts with a written description of a learning goal, uses the crowd to suggest predictive features, and then weighs these features using ML to produce models that are accurate and use human-understandable features. Specifically, Flock uses comparative examples to help crowd workers better nominate subtle features, and then adopts a clustering approach to re-organize these features. Flock is shown to be more effective than pure ML approaches. In the work of [Deng 2016], the authors introduced a novel approach for fine-grained image classification tasks (e.g., distinguish between "Northern Flicker" and "Red Bel-lied Woodpecker"). The authors proposed an online crowdsourcing game that elicits discriminative features by crowd worker (player). During the game, the player can choose to reveal details of circular regions (called "bubbles"), with certain incentive mechanism. Additionally, crowd workers can help discover features for unsupervised learning. For example, the authors in [Zou 2015] introduce an unsupervised approach to effectively discover features in a dataset via photo crowdsourcing. This approach first asks crowd workers to articulate a feature that is common to two out of three displayed photos and then requires the crowd workers to provide binary labels to the remaining photos based on the discovered features. It adaptively repeats the above process based on the labels of the previously discovered features.

**5.2 Learning Model Selection and Steering**

In additional to constructing a set of discriminative features, selecting appropriate models (e.g., the adopted classifiers in classification tasks) are also important for the performance of ML. Interactive model steering can help people incrementally build machine learning models that are tailored to their domains and tasks. Studies such as [Das 2019] aim to help ML engineers to select the most suitable models for corresponding data and learning tasks, where human-computer-interaction (HCI) tools are built to enable the interactive comparison and selection of different models. Specifically, it proposed a technique to allow users to inspect and steer multiple machine learning models. The technique steers and samples models from a broader set of learning algorithms and model types. More extensive review can be found in the survey [Riccardo 2019].

**5.3 Crowd Mining over Crowdsourced Data**

With the development of mobile social networks and mobile crowd sensing (MCS), more and more crowdsourced data are generated on the Web or collected from real-world sensing. The fragmented, heterogeneous, and noisy nature of online/offline crowdsourced data, however, makes it difficult to be understood. Traditional content-based analysis methods suffer from issues such as computational intensiveness and poor performance. To address this, a new research direction of HML, entitled CrowdMining [Guo 2019], has emerged. Based on the observation that the knowledge hidden in the process of data generation, including individual/crowd behavior patterns (e.g., mobility patterns, community contexts such as social ties and structure) and crowd-object interaction patterns (flickering or tweeting patterns), are neglected in crowdsourced data mining, it advocates to leverage the so-called '*crowd intelligence*' for crowdsourced data mining and understanding.

Various types of crowd intelligence are embedded in the data sensing or content generation process. These include interaction contexts based on human perception, decision making and opinions, and community contexts related to individual traits, community structure, and social/individual behavior patterns. They are often used indirectly for crowdsourced data understanding, i.e., used as features or parameter inputs of MI. There are various data mining tasks that can be performed by means of crowdsourcing, such as data filtering, classification, and clustering. For example, FlierMeet [Guo 2014] is an MCS app for public information (distributed fliers in the city) sensing and tagging. It identifies a novel set of crowd-object interaction hints (e.g., crowd-object interaction entropy, temporal patterns of interaction, interaction frequency) to predict the semantic tags of crowdsensed flier pictures.

**6. Interactive Model Assessment & Explanation**

In addition to data preparing and model learning phases, human can also be heavily involved in the model evaluation and explanation process.

**6.1 Interactive Quality/Robustness Assessment**

The training and refinement of ML models is an interactive process. The goal of AI engineers is to train the model to some level of acceptable accuracy. Therefore, the quality of ML models should be appropriately evaluated to determine if the model learning can be stopped. Amershi et al. [Saleema 2014] demonstrate visualization of the current model accuracy as an interface element present during the standard model steering task. In their work, the authors suggest that this visualization allows users to perform different strategies and evaluate their impact on accuracy. The quality assessment task may be interactive such as in the method proposed by [Saleema 2011], which provides a working environment where users can interact with the confusion matrix and express preferred classification requirements. In additional to the traditional quality evaluation on models, researchers recently started to focus on the robustness and safety of ML and AI systems (especially for deep-learning-based models), where a

ML system is interactively tested by the cooperation of human and machine under deliberately designed adversary cases [Ruan 2019].

**6.2 Model Explanation/Interpretation**

Understanding the reasons behind the suggestions (e.g., predictions or recommendations) made by AI systems is important in assessing trust, which helps end users to make final decisions (e.g., take an action or not). Understanding how a model works can also help refine an untrustworthy model into a trustworthy one.

A number of research works has started to study how users can evaluate and refine trained models when learning-based AI systems make incorrect decisions (e.g., classification, prediction, and recommendation). There are several methods or frameworks proposed in recent years to interpret ML models. Ribro et al. [Tulio 2016] developed a framework called LIME, which is able to explain predictions by learning an interpretable model locally around the prediction. For example, a model predicts that a patient has ae flu, and LIME highlights the symptoms in the patient's history that led to the prediction. Sneeze and headache are portrayed as contributing to the "flu" prediction, while "no fatigue" is evidence against it. With these, a doctor can make an informed decision about whether or not to trust the model's prediction. The authors in [Teso 2018] proposed the novel framework of explanatory interactive learning. In each step, the learner explains its interactive query to the user, and queries any active classifier for visualizing explanations of the corresponding predictions. In their study, the authors demonstrate that this framework can boost the predictive and explanatory powers of and the trust into the learned model.

Deep Learning models are powerful but being criticized the least interpretable. To this end, progress in this field have been accomplished by producing interpretation of the features learned at each layer of a Neural Network [Yosinski 2015] [Olah 2018]. A more extensive literature review about interpretability in deep learning can be found in [Zhang 2018].

**7. Opportunities and Proposals**

While various techniques have been proposed to improve ML with the human in the loop, several challenges still remain. In this section, we will highlight some existing gaps and propose new visions that can stimulate the research community to pursue new directions (i.e., new research problems).

**P1: Human-Machine Collaboration Fundamentals**

Although there are increasing number of technical researches on human-machine collaboration in the context of ML, the fundamental collaboration model and mechanism is still unclear. For example, what is the collaboration mode for human and machine? For the parallel mode, how will the output of machine and human (e.g., machine-extracted and human-nominated features) be integrated? For the sequential mode, how will two sides handoff to each other on different learning tasks. Besides, for different tasks, who are the human collaborators? Ordinary citizens or someone with specialized expertise? How many human workers should be involved? By successfully answering these questions, we can derive a clear picture on the model and mechanism of how human and machine is cooperated in the context of ML.

**P2: Multi-Objective Optimization**

When introducing human into the multiple tasks of the ML process, we have to consider multiple factors to optimize the utilization process itself. For example, from the perspective of the human workers, we have to consider factors such as intrusiveness, motivations, and their expertise/reputation. From the perspective the ML-related task requestors, task quality, budget, completion time, etc. In general, we formulate such research problem as an optimization with ML-task-specific goals and constraints. For example, a novel research problem can be how to select a minimum number of human workers to nominate features while ensuring a certain level of model prediction accuracy before the required deadline. Specifically, we need to further characterize and model the above factors in the context of ML applications and understand the difference compared with other types of crowdsourcing/crowdsensing tasks (e.g., urban environment sensing, software development, etc.)

**P3: Bias and Ethics Issue**

Task requesters usually overlook human workers behind tasks, this can lead to issues of ethics (e.g., unfair payment) and amplification of human biases, which are transferred into training data and affect machine learning in the real world. Research works such as [Barbosa 2019] have started to look into these issues and proposed feasible solutions, but most of them only focus on the crowdsourcing labeling tasks, while ignoring other tasks in the lifecycle of HML. Besides, we still need to understand human biases in HML, and interesting research topics could consist of effect of cultural, gender and ethnic biases, effect of human in the loop training and past experiences, effect of human expertise vs interest, bias in experts vs bias in crowdsourcing, bias in task selection, and task assignment/recommendation for reducing bias.

**P4: Integration of Human-Machine-Hybrid Features**

Human intelligence can help generate human-understandable features, and further by combining them with machine-extracted features, we have the potential to build a more powerful learning model [Cheng 2015]. However, there are two key critical remaining challenges that existing works do not address in this step. *First, how to obtain the value of human-nominated features?* As many of these features are subtle, machines cannot automatically assign values to them. Instead, humans have to handle many of these labeling tasks. This feature value acquisition task is far more costly than the classification labeling task, because for a certain training record (say, for a classification task), there is only one classification label but with numerous features. Therefore, it is crucial to study how to obtain the value of crowd-nominated features with the budget concern in mind. One promising research direction is to study the selection of most informative features with the objective of achieving a good tradeoff between the total feature labeling cost and model classification/prediction performance. *Second, how to integrate human and machine features?* When combining the human-generated features with machine-extracted ones, a naïve approach is to directly connect these two types of features as a longer feature vector. However, an alternative approach could be training machine and human classifiers separately, then combining them by adopting ensemble learning techniques to make a more accurate prediction.

**P5: From Individual to Crowd**

Most existing studies in HML focus on the interaction of one single human worker with a ML system. With the popularity of social networks (e.g., twitter, Facebook, etc.) and crowdsourcing platforms in different domains (e.g., product design, software engineering, etc.), there is now an opportunity of recruiting large number of crowd workers collaboratively in a ML system. Therefore, techniques and tools should be developed to support the collaborative work in the context of building ML models or applications. For instance, a collaborative working environment should be developed to help workers to discuss, comment, share, and nominate more efficiently. We have already witnessed substantial progress in crowdsourcing-based labeling tasks. Nevertheless, further research on how the crowd can work collaboratively in other higher-level tasks such as feature selection, evaluation, and interpretations of models, is still needed.

## 8. Conclusion

This paper presents a review of ML with human in the loop. Specifically, by organizing existing studies along the multi-stage lifecycle of ML, we present various kinds of strategies exploiting human intelligence in ML. in addition, we point introduce some research gaps and directions that may further integrate the power of human and machine intelligence to build more cost-effective, reliable, and robust intelligent systems.

# References

[Barbosa 2019] Barbosa, et al. "Rehumanized Crowdsourcing: A Labeling Framework Addressing Bias and Ethics in Machine Learning." CHI 2019, p. 543. ACM, 2019.

[Cheng 2015] Cheng, J., et al. (2015, February). Flock: Hybrid crowd-machine learning classifiers. In Proceedings of the 18th ACM Conference on Computer Supported Cooperative Work & Social Computing (pp. 600-611).

[Chu 2015] Chu, X., et al. (2015, May). Katara: A data cleaning system powered by knowledge bases and crowdsourcing. In Proceedings of the 2015 ACM SIGMOD International Conference on Management of Data (pp. 1247-1261).

[Das 2019] Das, Subhajit, et al. "BEAMES: Interactive Multi-Model Steering, Selection, and Inspection for Regression Tasks." IEEE computer graphics and applications (2019).

[Deng 2016] Deng, J., et al. (2016). Leveraging the wisdom of the crowd for fine-grained recognition. IEEE transactions on pattern analysis and machine intelligence, 38(4), 666-676.

[Dudley 2018] Dudley, et al. "A review of user interface design for interactive machine learning." ACM Transactions on Interactive Intelligent Systems, 8, no. 2 (2018): 8.

[Guo 2014] Guo, Bin, et al. "FlierMeet: a mobile crowdsensing system for cross-space public information reposting, tagging, and sharing." IEEE Transactions on Mobile Computing 14.10 (2014): 2020-2033.

[Guo 2019] Guo, Bin, et al. "From crowdsourcing to crowdmining: using implicit human intelligence for better understanding of crowdsourced data." World Wide Web (2019): 1-25.

[Kamar 2016] Kamar E. Directions in Hybrid Intelligence: Complementing AI Systems with Human Intelligence. In IJCAI 2016, Jul 9, pp. 4070-4073.

[Olah 2018] C. Olah, et al. The Building Blocks of Interpretability. Distill, 2018. doi: 10.23915/distill.00010

[Li 2017] Li G, et al. Crowdsourced data management: Overview and challenges. In Proceedings of the 2017 ACM International Conference on Management of Data, 2017 May 9 (pp. 1711-1716). ACM.

[Liu 2018] Yang Liu et al. Doubly Active Learning: when Active Learning meets Active Crowdsourcing. AAAI 2018.

[Howe 2006] Howe J. The rise of crowdsourcing. Wired magazine. 2006 Jun 6;14(6):1-4.

[Nushi 2015] Nushi, B., et al. (2015, September). Crowd access path optimization: Diversity matters. In Third AAAI Conference on Human Computation and Crowdsourcing.

[Riccardo 2019] Guidotti, Riccardo, et al. "A survey of methods for explaining black box models." ACM computing surveys (CSUR) 51, no. 5 (2019): 93.

[Saleema 2014] Amershi, Saleema, et al.. "Power to the people: The role of humans in interactive machine learning." AI Magazine 35, no. 4 (2014): 105-120.

[Saleema 2011] Amershi, Saleema, et al.. "Effective end-user interaction with machine learning." In Twenty-Fifth AAAI Conference on Artificial Intelligence. 2011.

[Ruan 2019] Ruan, Wenjie, et al. "Global Robustness Evaluation of Deep Neural Networks with Provable Guarantees for the Hamming Distance." IJCAI, 2019.

[Teso 2018] S. Teso and K. Kersting. "Why Should I Trust Interactive Learners?" Explaining Interactive Queries of Classifiers to Users. arXiv preprint, 5 2018.

[Tong 2014] Tong, Y., et al. (2014, March). Crowdcleaner: Data cleaning for multi-version data on the web via crowdsourcing. In Data Engineering (ICDE), 2014 IEEE 30th International Conference on (pp. 1182-1185). IEEE.

[Tulio 2016] Ribeiro, et al.. "Why should I trust you?: Explaining the predictions of any classifier." In Proceedings of the 22nd ACM SIGKDD international conference on knowledge discovery and data mining, pp. 1135-1144. ACM, 2016.

[Venanzi 2016] Venanzi, et al. (2016). Time-sensitive bayesian information aggregation for crowdsourcing systems. Journal of Artificial Intelligence Research, 56, 517-545.

[Wang 2019] Wang J, et al.. Crowd-Assisted Machine Learning: Current Issues and Future Directions. Computer. 2019 Mar 13;52(1):46-53.

[Xu 2016] Chu, Xu, et al.. "Data cleaning: Overview and emerging challenges." In Proceedings of the 2016 international conference on Management of Data, pp. 2201-2206. ACM, 2016.

[Yosinski 2015] J. Yosinski, et al. Understanding Neural Networks Through Deep Visualization. In ICML Workshop on Deep Learning, 2015.

[Zhang 2018] Zhang C, et al. Reducing uncertainty of schema matching via crowdsourcing with accuracy rates. IEEE Transactions on Knowledge and Data Engineering. 2018.

[Zhang 2018] Zhang QS, Zhu SC. Visual interpretability for deep learning: a survey. Frontiers of Information Technology & Electronic Engineering. 2018 Jan 1;19(1):27-39.

[Zhang 2019] Ruohan Zhang, et al.. Leveraging Human Guidance for Deep Reinforcement Learning Tasks. In Proceedings of the 28th International Joint Conference on Artificial Intelligence (IJCAI), August 2019.

[Zhou 2017] Zhou ZH. A brief introduction to weakly supervised learning. National Science Review. 2017 Aug 25;5(1):44-53.

[Zhuang 2017] Zhuang Y, et al. Hike: A hybrid human-machine method for entity alignment in large-scale knowledge bases. In Proceedings of the 2017 ACM on Conference on Information and Knowledge Management 2017 Nov 6 (pp. 1917-1926). ACM.

[Zou 2015] Zou, J. Y., et al. (2015). Crowdsourcing feature discovery via adaptively chosen comparisons. Proceedings of the 31st International Conference on Machine Learning, Lille, France